\documentclass{article}
 \usepackage[dvipdf]{epsfig}
  \usepackage{color}
  \usepackage{amssymb}
\begin{document}
\newcommand{\ben}{\begin{eqnarray}}
\newcommand{\een}{\end{eqnarray}}

%
%

\title{The van der Waals fluid and its role in cosmology}

\author{Rudinei C. S. Jantsch$^1$\footnote{rudijantsch@gmail.com}, Marcus H. B. Christmann$^2$\footnote{marcushbc@gmail.com} and Gilberto M. Kremer$^2$\footnote{kremer@fisica.ufpr.br}\\
$^1$Instituto Federal Sul-rio-grandense\\
Santana do Livramento, Rio Grande do Sul 97574-360, Brazil\\
$^2$Departamento de F\'\i sica, Universidade Federal do Paran\'a\\
Curitiba, Paran\'a 81531-980, Brazil
}

\date{}
\maketitle

\begin{abstract}
We analyze the properties of a generic cosmological fluid described by the van der Waals equation of state. Exact solutions for the energy density evolution
are found as implicit functions of the scale factor for a flat Friedmann-Robertson-Walker space-time. The possible values of the free parameter in the
van der Waals equation are selected \emph{a posteriori}, in accordance with asymptotic behaviors that are physically relevant. The stability of the model
against small perturbations is studied through the hydrodynamic perturbations of the fluid for the relevant cases. It is found that a van der Waals fluid seems
appropriate to describe non-eternal inflationary scenarios.
\end{abstract}


\section{Introduction}

With the advent of the inflationary cosmology \cite{guth, linde} and the relatively recent discovery of the accelerated expansion of the Universe at late times
\cite{riess, perlmutter}, the cosmological components enlarged and exotic equations of state were needed to model the acceleration of the cosmic expansion
\cite{caldwell, steinhardt, peebles}. The Chaplygin gas type equation of state is frequently studied in the literature and used to describe standard matter and
dark energy as a single fluid \cite{chaplygin0, chaplygin0.1, chaplygin0.2, chaplygin1, chaplygin2, chaplygin3, chaplygin4}. Such equations of state can be
generated by a standard scalar field with a specific potential \cite{chaplygin0, chaplygin5} or a tachyon scalar field \cite{chaplygin6}, which can be also
applied in various other contexts
\cite{tachyon1, tachyon2, tachyon3}.

In the concept of standard matter and dark energy as a single fluid, the van der Waals equation of state has received some attention and was investigated in the literature \cite{vanderwaals1, vanderwaals1.1, vanderwaals1.2}.
The van der Waals fluid as a cosmological constituent in a variety of other contexts was also investigated
\cite{vanderwaals2, vanderwaals2.1, vanderwaals3, vanderwaals4, vanderwaals5}.
Due to the non-linearities of such equations of state, the cosmological dynamics is usually solved numerically and asymptotic analytic solutions can be obtained
just in some special cases. The main feature explored in these models concerns the ability to reproduce the accelerated and matter-dominated ages with a single
component. Further, depending on the characteristics of the model in question, the properties of the early and/or the late Universe may be possible to describe.

In this work, we do not impose any desired behavior on the van der Waals fluid. Instead we look for general analytical solutions of the energy density in a
flat Friedmann-Robertson-Walker space-time, and so select the physically relevant cases. Once the suitable cases are selected, the corresponding stability is
tested through the analysis of the hydrodynamic perturbations.

\section{Basic equations}

Let us consider a cosmological fluid that obeys the van der Waals equation of state, which is written in the following form \cite{vanderwaals2}
\begin{equation}
p=\frac{8w\rho}{3-\rho}-3\rho^2, \label{vweq}
\end{equation}
where $p$ and $\rho$ are the pressure and energy density of the fluid, respectively. The parameter $w$ is a non-negative constant. This non-linear equation of
state presents some peculiar qualitative features. It has a vertical asymptote in $\rho=3$, where the pressure diverges, and its form approaches the form
$p\propto \rho$ when $\rho$ is sufficiently small while the pressure is persistently negative when $\rho$ is large enough. Further, the pressure can eventually
switch from a positive value to a negative one and vice versa. Some of these features are very compelling for cosmology and justify a more detailed investigation
of the van der Waals fluid in this context.

In a flat Friedmann-Robertson-Walker (F-R-W) background, assuming that the perfect fluid conservation equation holds, we have
\begin{equation}
\dot{\rho}+3\frac{\dot a}{a}\left(\rho+p\right)=0, \label{conseq}
\end{equation}
with the dot denoting time derivative and $a$ the scale factor.

From equations (\ref{vweq}) and (\ref{conseq}), one obtains a differential equation for $a$ and $\rho$, namely
\begin{equation}
a(3-\rho)\dot{\rho}+3\dot a\rho\left(3+8w-10\rho+3\rho^2\right)=0,
\end{equation}
which can be rewritten in the form
\begin{equation}
a(3-\rho)\frac{d\rho}{da}+3\rho\left(3+8w-10\rho+3\rho^2\right)=0.
\label{diffeq}
\end{equation}
This differential equation furnishes the necessary information to determine the energy density evolution of the fluid.

By assuming that general relativity governs the large scale dynamics, the energy density of the van der Waals fluid has an effect on the cosmic evolution
through the Friedmann equation
\begin{equation}
H^2=\left(\frac{\dot a}{a}\right)^2=\frac{\rho}{3},
\label{FE}
\end{equation}
where $H$ stands for the Hubble parameter.

\section{Background solutions}

With the above equations in hand, we can now try to obtain the functional behavior of the energy density of the van der Waals fluid in a flat F-R-W background.
Equation (\ref{diffeq}) is separable in terms of functions of $a$ and $\rho$, generating the following differential expression
\begin{equation}
\frac{da}{a}=\frac{1}{3}\ \frac{d\rho}{3+8w-10\rho+3\rho^2}-\frac{1}{\rho}\ \frac{d\rho}{3+8w-10\rho+3\rho^2}.
\end{equation}
The corresponding integration furnishes
\begin{eqnarray}
\ln{a}=\frac{1}{2(3+8w)}\ln{\left|\frac{3+8w-10\rho+3\rho^2}{\rho^2}\right|}+\frac{4(2w-3)}{3(3+8w)}\int\frac{d\rho}{3+8w-10\rho+3\rho^2}+k,\nonumber\\
\label{diffeq1}
\end{eqnarray}
where $k$ is a constant of integration. The integral in the right hand side of the above expression contains a polynomial of the form $A\rho^2+B\rho+C$, with $A=3$, $B=-10$ and $C=3+8w$,
whose integration for $4AC-B^2>0$ gives
\begin{equation}
a=\frac{C_0}{\rho^{\frac{1}{3+8w}}}\left(3+8w-10\rho+3\rho^2\right)^{\frac{1}{2(3+8w)}}
\exp{\left(\frac{2(2w-3)\arctan\left(\frac{3\rho-5}{2\sqrt{6w-4}}\right)}{3(3+8w)\sqrt{6w-4}}\right)}.
\label{sol1}
\end{equation}
Here $C_0$ is a constant and this solution is valid for $w>2/3$.

In the case for $4AC-B^2=0$, the solution is
\begin{equation}
a=\frac{C_0}{\rho^{\frac{1}{3+8w}}}\left(3+8w-10\rho+3\rho^2\right)^{\frac{1}{2(3+8w)}}\exp{\left(\frac{4(2w-3)}{3(3+8w)(5-3\rho)}\right)}\,
\label{sol2}
\end{equation}
which is valid only for $w=2/3$.

Finally, the solution corresponding to the case $4AC-B^2<0$ reads
\begin{equation}
a=\frac{C_0}{\rho^{\frac{1}{3+8w}}}\left(3+8w-10\rho+3\rho^2\right)^{\frac{1}{2(3+8w)}}\left(\frac{3\rho-5-2\sqrt{4-6w}}{3\rho-5+2\sqrt{4-6w}}\right)
^{\frac{(2w-3)}{3(3+8w)\sqrt{4-6w}}},\label{sol3}
\end{equation}
which holds for $0<w<2/3$.

All the above solutions have asymptotic roots, i.e. they allow values of $\rho$ that satisfy $a\longrightarrow0$, meaning that the equation of state considered
here is compatible with what is expected for the Universe at early times. As we can realize, expressions (\ref{sol1}), (\ref{sol2}) and
(\ref{sol3}) involve terms quadratic in $\rho$ and one can expect two roots for each of them. However, there is only one root common to all solutions and another
is a root only of (\ref{sol1}) and (\ref{sol2}). Let us see this in more details as follows.

Expressions (\ref{sol1}), (\ref{sol2}) and
(\ref{sol3}) satisfy
\begin{equation}
\rho\longrightarrow \rho_0=(5-2\sqrt{4-6w})/3, \quad \textrm{when} \quad a\longrightarrow 0.\label{root}
\end{equation}
So $\rho_0$ is an asymptotic root for all the above solutions. But notice that for $w>2/3$, solution (\ref{sol1}), we have that (\ref{root}) is complex and must
be discarded for physical purposes. Then the value given by (\ref{root}) is physically valid for $0<w\leq2/3$ and is the asymptotic initial condition from which
the Universe expands as $\rho$ decreases. Here it is important to point out that such a property means that a van der Waals fluid does not have a divergent
energy density when $a\longrightarrow 0$, i.e. it does not present a big bang singularity.
Interestingly enough, is the result from equation (\ref{vweq}) when evaluated at $\rho=\rho_0=(5-2\sqrt{4-6w})/3$, namely,
\begin{equation}
p=-\rho_0.\label{vacuum_energy}
\end{equation}
This result establishes that for a sufficiently small $a$ the van der Waals dynamics for $0<w\leq2/3$, encompassed by (\ref{sol2}) and (\ref{sol3}), is close to
the one of the cosmological constant. Hence, the equation of state $p\approx-\rho$ holds in the beginning of the expansion and we have a de Sitter-like inflation
for $0<w\leq2/3$. One more important point to stress is that once we take initial conditions around (\ref{root}), the energy density will never take values
near the vertical asymptote, $\rho=3$, given that $\rho_0<3$ and $\rho$ always will decrease with the increase of $a$.

The second asymptotic root is
\begin{equation}
\rho\longrightarrow \rho_0'=(5+2\sqrt{4-6w})/3, \quad \textrm{when} \quad a\longrightarrow 0,\label{root'}
\end{equation}
which holds only for (\ref{sol1}) and (\ref{sol2}) - solution (\ref{sol3}) diverges for $\rho\longrightarrow\rho_0'$. This root also produces a complex $\rho$
for $w>2/3$, but even for $w=2/3$, solution (\ref{sol2}), it is not physically motivated as an asymptotic initial condition because it leads to limiting
equations of state that are not viable.

Now consider the dynamics in the range $0<w\leq2/3$, were the initial condition (\ref{root}) holds and the energy density dilutes as the Universe expands. Then,
from solutions (\ref{sol2}) and (\ref{sol3}) it is straightforward to see that when $\rho\longrightarrow 0$,  i.e. $a$ is large, we have
\begin{equation}
\rho\propto a^{-(3+8w)} \quad \Longrightarrow \quad a\propto t^{\frac{2}{3+8w}}.\label{limitofa}
\end{equation}
This is in accordance with (\ref{vweq}), from which results $p=8w\rho/3$ in the referred limit.

As follows, we will analyze the cases of physical interest: $w=0$ and $w=1/8$.\\

1) Case $w=0$

It corresponds to the equation of state $p=-3\rho^2$, see equation (\ref{vweq}), which through (\ref{conseq}) furnishes the differential expression
\begin{equation}
3\frac{da}{a}=\frac{d\rho}{3\rho^2-\rho}.
\end{equation}
The respective integration renders
\begin{equation}
\rho=(c_0a^3+3)^{-1}, \label{sol4}
\end{equation}
where $c_0$ is a constant of integration.

By taking the Friedmann equation, $H^2=\rho/3$, and applying (\ref{sol4}), we have the differential expression bellow
\begin{equation}
\sqrt{c_0a^3+3}\ \frac{da}{a}=\frac{dt}{\sqrt{3}} \quad \Longrightarrow \quad 2\sqrt{c_0u^2+3}\ \frac{du}{u}={\sqrt{3}}\ dt, \label{diffscale}
\end{equation}
where the change of variable $u^2=a^3$ was performed.

Integration of (\ref{diffscale}) gives the implicit time evolution of the scale factor, namely,
\begin{equation}
\sqrt{c_0a^3+3}+\sqrt{3}\ln{\left(\frac{a^{3/2}}{6+2\sqrt{3(c_0a^3+3)}}\right)}=\frac{\sqrt{3}}{2}t+C_1, \label{scalef}
\end{equation}
with $C_1$ being a constant of integration.

The above expression has the asymptotic behaviors
\begin{equation}
a\propto \exp{\left(\frac{t}{3}\right)}, \label{scalef1}
\end{equation}
for a sufficiently small $a$, and
\begin{equation}
a\propto t^{2/3}, \label{scalef2}
\end{equation}
corresponding to a large $a$. Hence, this case describes an inflation followed by a matter-dominated era type dynamics.\\

2) Case $w=1/8$

The range of validity of solution (\ref{sol3}) comprises $w=1/8$, from which we have
\begin{equation}
a=\frac{C_0}{\rho^{1/4}}\left(4-10\rho+3\rho^2\right)^{1/8}\left(\frac{3\rho-5+\sqrt{13}}{3\rho-5-\sqrt{13}}\right)^
{\frac{11}{24\sqrt{13}}}.\label{sol3.1}
\end{equation}
From the general result (\ref{root}), this expression has the root $\rho\longrightarrow \rho_0=(5-\sqrt{13})/3$, which is the initial energy density and
corresponds
to the equation of state $p=-\rho_0$, as stated by (\ref{vacuum_energy}).
Also from (\ref{sol3.1}), when $\rho$ is sufficiently diluted we have $\rho\propto a^{-4}$, i.e. a radiation-dominated era, as anticipated in (\ref{limitofa}).
Thus this case describes an inflationary period followed by a radiation-dominated era.

\section{Hydrodynamic perturbations}

In this section we will develop the general equations to evaluate the small scalar perturbations in a flat F-R-W background for a van der Waals fluid.
As the background fluid does not present anisotropy, the line element for small scalar perturbations in the longitudinal gauge is
\cite{bardeen, mukhanov, mukhanov1}
\begin{equation}
ds^2=a^2\left[\left(1+2\Phi\right)d\eta^2-\left(1-2\Phi\right)\delta_{ij}dx^idx^j\right],
\end{equation}
where $\eta$ is the conformal time, $d\eta=dt/a$, and $\Phi$ is the gauge-invariant perturbation potential.

By introducing the Mukhanov-Sasaki variables \cite{mukhanov, mukhanov1}, which are defined by
 \begin{equation}
\sigma=\frac{\Phi}{\sqrt{\rho+p}}, \qquad \chi=\left(\frac{1}{a}\right)\left(1+\frac{p}{\rho}\right)^{-1/2},\label{ms_variables}
\end{equation}
the adiabatic perturbations, in the general relativity context, are described by the following compact differential equation
 \begin{equation}
\sigma''-v_s^2\nabla^2\sigma-\left(\frac{\chi''}{\chi}\right)\sigma=0,\label{perteq}
\end{equation}
where the prime represents derivative with respect to the conformal time and $v_s^2=\partial p/\partial \rho$ is the speed of sound.

For a plane-wave solution of the form $\sigma(\eta)e^{-i\textbf{k.x}}$, the \emph{short-wavelength perturbations} hold the condition $v_sk\eta\gg 1$ and the
last term of equation (\ref{perteq}) can be neglected, furnishing
\begin{equation}
\sigma''+k^2v_s^2\sigma=0.\label{perteq1}
\end{equation}

The corresponding solution, valid for a sufficient slowly varying speed of sound (WKB approximation) reads
\begin{equation}
\sigma(\eta)\simeq \frac{\sigma_0}{\sqrt{v_s}}\exp\left({\pm ik\int{v_sd\eta}}\right),\label{SOLA}
\end{equation}
where $\sigma_0$ is a constant of integration.

For \emph{long-wavelength perturbations}, when the condition $v_sk\eta\ll1$ is valid, the spatial derivatives of (\ref{perteq}) can be neglected and the
following solution holds
\begin{equation}
\sigma(\eta)\simeq \sigma_0\chi\int\frac{d\eta}{\chi^2}.\label{SOLB}
\end{equation}

All the quantities appearing in the asymptotic solutions above can be expressed in a compact form as functions of the energy density $\rho$.
The Friedmann equation furnishes the infinitesimal conformal time interval in the form
\begin{equation}
d\eta=\frac{dt}{a}=\sqrt{3}\frac{da}{a^2\sqrt{\rho}}\ .\label{conformal}
\end{equation}
On the other hand, by using equation of state (\ref{vweq}) one directly determines the speed of sound, namely,
\begin{equation}
v_s=\sqrt{\frac{\partial p}{\partial\rho}}=\sqrt{\frac{24w}{(3-\rho)^2}-6\rho}\ .\label{speedofsound}
\end{equation}
Also equation (\ref{vweq}) allows to write the variable $\chi$ given by $(\ref{ms_variables})_2$ as a function of $\rho$ in the form
\begin{equation}
\chi=\frac{1}{a}\left[{\frac{3+8w-10\rho+3\rho^2}{3-\rho}}\right]^{-1/2} .\label{chivariable}
\end{equation}

Furthermore, the variable $\chi$ is also equivalent to
 \begin{equation}
\chi=\frac{1}{a}\left[\frac{2}{3}\left(1-\frac{\mathcal{H}'}{\mathcal{H}^2}\right)\right]^{-1/2},\label{chivariable1}
\end{equation}
where $\mathcal{H}=a'/a$ is the Hubble parameter in terms of the conformal time. To get (\ref{chivariable1}) from (\ref{chivariable}) one uses the acceleration
equation
 \begin{equation}
\frac{\mathcal{H}'}{a^2}=-\frac{1}{6}\left(\rho+3p\right),
\end{equation}
along with the equation of state (\ref{vweq}).
From the variable $\chi$ in the form (\ref{chivariable1}) one can write the integral in (\ref{SOLB}) as
 \begin{equation}
\int\frac{d\eta}{\chi^2}=\frac{2}{3}\left(\frac{a^2}{\mathcal{H}^2}-\int a^2d\eta\right),
\end{equation}
through an integration by parts. Using this result in (\ref{SOLB}) and returning to $(\ref{ms_variables})_1$, with a plane-wave solution of
the form $\Phi(\eta)e^{-i\textbf{k.x}}$, finally we have
 \ben
 &\Phi(\eta)=\sigma_1\left(1-\frac{\mathcal{H}}{a^2}\int a^2d\eta\right)=\sigma_1\left(1-\frac{1}{a}\sqrt{\frac{\rho}{3}}\int a^2d\eta\right)\nonumber\\
 &\Phi(\eta)=\sigma_1\left(1-\frac{\sqrt{\rho}}{a}\int\frac{da}{\sqrt{\rho}}\right),\label{pertpotential}
\een
where (\ref{conformal}) was used and $\sigma_1$ is a constant defined by $\sigma_1 = 2\sigma_0/\sqrt{3}$. With the quantities (\ref{SOLA}),
(\ref{conformal}), (\ref{speedofsound}), (\ref{chivariable}) and (\ref{pertpotential}) in hand, we can evaluate the perturbations for any $w$ and determine the
density contrast. The perturbed dynamics for the cases of interest will be analyzed as follows.

Let us now apply the above results to the case 1, when $w=0$, and case 2, when $w=1/8$. We immediately infer from equation of state of the case 1, $p=-3\rho^2$,
that the speed of sound (\ref{speedofsound}) is always imaginary, which will generate perturbations with increasing amplitude all the time. Hence, the case 1 is
unstable against small perturbations. The same occurs for the case 2 when $a$ is small, since in this situation $p\approx -\rho$ and solution (\ref{SOLA}) will
give a combination of real exponentials for short-wavelength perturbations. However, the equation of state for the case 2 rapidly evolves to $p=\rho/3$
(radiation) as $\rho$ dilutes with inflation, so reaching a real speed of sound and it is expected that persistent growing instabilities are prevented. Thus from
now on we are focused on the perturbations for the case with $w=1/8$, which is in fact the most interesting one since it naturally describes inflation followed
by a radiation-dominated era.

 Substituting (\ref{sol3.1}) into (\ref{pertpotential}), we get the perturbation potential for $w=1/8$ in the long-wavelength limit as a function
 only of $\rho$, namely,
 \begin{equation}
\Phi(\rho)=1-\rho^{3/4}\left(4-10\rho+3\rho^2\right)^{-1/8}\left(\frac{3\rho-5-\sqrt{13}}{3\rho-5+\sqrt{13}}\right)^{\frac{11}{24\sqrt{13}}}\int_{\rho_0} f(\rho)d\rho, \label{pertpotentialcase2}
\end{equation}
where
 \begin{equation}
f(\rho)\doteq\frac{1}{\sqrt{\rho}}\frac{da}{d\rho}=\frac{(\rho-3)\left(\frac{3\rho-5+\sqrt{13}}{3 \rho-5-\sqrt{13}}\right)^{\frac{11}{24{\sqrt{13}}}}}
{3\rho^{7/4}(4-10\rho + 3\rho^2)^{7/8}},
 \label{rhofunction}
\end{equation}
and $\rho_0$ denotes the initial energy density corresponding to $a_0$. Here the constants $C_0$ and $\sigma_1$ are set to unity without loss of generality.

With this result in hand, using the following perturbation equation from the perturbed Einstein's equations
 \begin{equation}
 \nabla^2\Phi-3\mathcal{H}\left(\Phi'+\mathcal{H}\Phi\right)=\frac{a^2}{2}\delta\rho,\label{firstpert}
 \end{equation}
where $\delta\rho$ is the gauge-invariant density perturbation, we can now infer the density contrast.

Since we are analyzing the long-wavelength limit, the first term of (\ref{firstpert}) can be neglected (spatial derivatives).
Further, using the Friedmann equation in terms of the conformal time, $3\mathcal{H}^2=a^2\rho$, equation (\ref{firstpert}) furnishes
 \begin{equation}
 \frac{\delta\rho}{\rho}=-2\left(\frac{\Phi'}{\mathcal{H}}+\Phi\right),\label{contrast}
 \end{equation}
which gives the density contrast if $\Phi$ is known.

If we derive (\ref{pertpotential}) with respect to conformal time, one has
 \begin{equation}
\Phi'(\eta)=\frac{1}{a^2}\left(2\mathcal{H}^2-\mathcal{H}'\right)\int a^2d\eta-\mathcal{H}.\label{Philine}
 \end{equation}
Now we use (\ref{chivariable1}) to eliminate $\mathcal{H'}$ in the last expression, which is then substituted into (\ref{contrast}).
The corresponding result can be expressed in terms of $\rho$ by means of (\ref{conformal}) and (\ref{chivariable}).
So, by considering the density perturbation of the form $\delta\rho(\eta)e^{-i\textbf{k.x}}$, one obtains
 \begin{equation}
 \frac{\delta\rho}{\rho}\ (\rho)=-3{\rho}^{3/4}\frac{\left(4 - 10\rho + 3\rho^2\right)^{7/8}}{(3 - \rho)}
  \left(\frac{3\rho-5-\sqrt{13}}{3\rho-5+\sqrt{13}}\right)^{\frac{11}{24\sqrt{13}}}\int_{\rho_0} f(\rho)d\rho.\label{finalcontrast}
 \end{equation}
Thus we have an expression that gives the density contrast for $w=1/8$ as a function only of $\rho$.

\begin{figure}[ht]
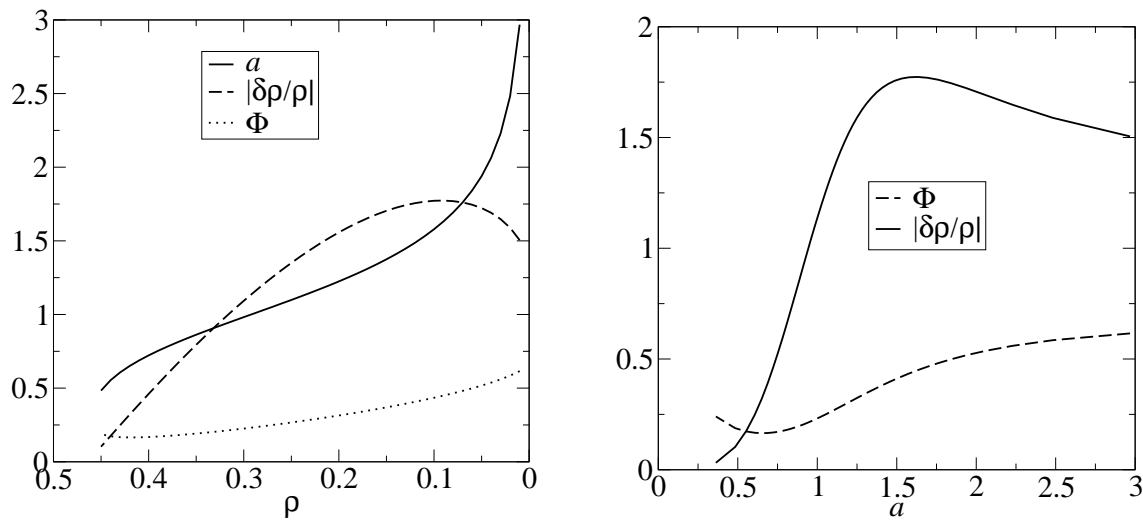
\vskip0.5cm
  \centering
  \includegraphics[width=0.4\textwidth]{fig2.eps}\hskip1cm\includegraphics[width=0.4\textwidth]{fig1.eps}
  \caption
  {Left: The amplitudes of the perturbation potential and density contrast as functions of the scale factor ($w=1/8$). Right: The scale factor and amplitudes of the density contrast and perturbation potential as functions of the energy density ($w=1/8$). }
  \label{fig1}
\end{figure}

The scale factor (\ref{sol3.1}) and amplitudes of the density contrast (\ref{finalcontrast}) and perturbation potential (\ref{pertpotentialcase2})
are plotted in the left frame of Fig.~\ref{fig1}. By observing the corresponding curves, we can see that the amplitude of the perturbation potential increases slowly with the dilution of $\rho$, i.e. with increasing $a$. On the other hand, the amplitude of the density contrast increases initially with the dilution of $\rho$ and starts to decrease when $\rho$ is sufficiently diluted, i.e. when $a$ is sufficiently large and a radiation-dominated era is expected.

An alternative visualization of the perturbations is shown in the right frame Fig.~\ref{fig1}, where the amplitudes of the perturbation potential (\ref{pertpotentialcase2})
and density contrast (\ref{finalcontrast}) are expressed as functions of the scale factor. From this figure it is clear to see the slow growth
of the amplitude of the perturbation potential with the cosmic expansion while the amplitude of the density contrast grows in the beginning and starts
to decrease after a certain point of the cosmic expansion.

In view of these results, the van der Waals fluid could describe inflation followed by a radiation era, i.e. a non-eternal inflation which is stable
against small perturbations in the limit of the long-wave perturbations, as shown by the above analysis of the hydrodynamic perturbations.

\section{Concluding summary}

In this work, we studied a generic cosmological fluid described by the van der Waals equation of state for a flat F-R-W space-time.
General background solutions for the energy density of the fluid were obtained in analytical form.
From these solutions it was possible to identify the physically relevant cases.
It was found that in general the energy density of a cosmological van der Waals fluid does not diverge as the scale factor goes to zero (no big bang singularity)
and behaves like a cosmological constant type component.
Two cases of interest in cosmology were found and they correspond to the values 0 and 1/8 for the parameter $w$ of the equation of state.
For $w=0$ the background solution describes an inflation followed by a matter-dominated era while for $w=1/8$ it describes an inflation followed
by a radiation-dominated era.

The viability of the relevant background solutions is tested through the analysis of the hydrodynamics perturbations.
The case for $w=0$ is unstable against small perturbations since the speed of sound is persistently imaginary.
On the other hand, the case for $w=1/8$ is stable against small perturbations in the limit of long-wave perturbations.
This entails that the van der Waals model presents some skill in describing a non-eternal inflation, i.e. an initial \emph{quasi}-cosmological constant dynamics
followed by a radiation-dominated era.

\section*{Acknowledgments}

M. H. B. C. acknowledges  the  Coordena\c c\~ao de Aperfei\c coamento de Pessoal de N\'ivel Superior (CAPES)  and G. M. K. the Conselho  Nacional
de Desenvolvimento Cient\'ifico e Tecnol\'ogico (CNPq), Brazil, for financial support.

\end{document}